\documentclass[twocolumn,showpacs,preprintnumbers,amsmath,amssymb,prl]{revtex4}

\usepackage{graphicx}
\usepackage{dcolumn}
\usepackage{bm}

\usepackage{mathptmx, courier, pifont}
\usepackage[scaled=0.92]{helvet}
\usepackage[T1]{fontenc}
\usepackage{textcomp}
\usepackage{color}
\usepackage{colordvi}

\newcommand{\quarterthin}{\kern 0.0417em}

\begin{document}


\title{Isospin nonconserving interaction in the $T=1$ analogue states
of the mass-$70$ region}

\author{K.~Kaneko$^{1}$, Y.~Sun$^{2,3}$, T.~Mizusaki$^{4}$, S.~Tazaki$^{5}$ }

\affiliation{
$^{1}$Department of Physics, Kyushu Sangyo University, Fukuoka
813-8503, Japan \\
$^{2}$Department of Physics and Astronomy, Shanghai Jiao Tong
University, Shanghai 200240, People's Republic of China \\
$^{3}$Institute of Modern Physics, Chinese Academy of Sciences,
Lanzhou 730000, People's Republic of China \\
$^{4}$Institute of Natural Sciences, Senshu University, Tokyo
101-8425, Japan \\
$^{5}$Department of Applied Physics, Fukuoka University, Fukuoka
814-0180, Japan
}

\date{\today}

\begin{abstract}

Mirror energy differences (MED) and triplet energy differences (TED)
in the $T=1$ analogue states are important probes of
isospin-symmetry breaking. Inspired by the recent spectroscopic data
of $^{66}$Se, we investigate these quantities for $A=66-78$ nuclei
with large-scale shell-model calculations. For the first time, we
find clear evidences suggesting that the isospin nonconserving (INC)
nuclear force has a significant effect for the upper $fp$ shell
region. Detailed analysis shows that in addition to the INC force,
the electromagnetic spin-orbit interaction plays an important role
for the large, negative MED in $A=66$ and 70 and the multipole
Coulomb term contributes to the negative TED in all the $T=1$
triplet nuclei. The INC force and its strength needed to reproduce
the experimental data are compared with those from the G-matrix
calculation using the modern charge-dependent nucleon-nucleon
forces.

\end{abstract}


\maketitle


The impact of the Wigner's elegant concept, the isospin symmetry
\cite{Wigner37}, is maximal near the $N=Z$ line where nuclei have
equal numbers of neutrons and protons. Breaking of this symmetry is
generally attributed to the Coulomb and isospin-nonconserving (INC)
nuclear forces. To study the isospin-symmetry breaking, information
for nuclei with $N<Z$ is of particular interest but these nuclei are
not easy to access experimentally. By comparison of nuclear masses
\cite{Zhang12} and detailed spectroscopic information
\cite{Warner06} for nuclei having same isospin, $T$, one can study
the isospin-related phenomena to explore the origin of the symmetry
breaking.

Measurable quantities have been suggested to probe the
isospin-symmetry breaking. Mirror energy differences (MED), which
are the differences between excitation energies of the $T=1$
isobaric analogue states (IAS), are regarded as measures of the
charge-symmetry breaking. On the other hand, triplet energy
differences (TED) among the triplet $T=1$ nuclei are used to
indicate the charge-independence breaking. MED were extensively
studied for the $f_{7/2}$-shell nuclei up to high spins (see Ref.
\cite{Bentley07} for review).  TED were discussed for the $A=46$
\cite{Garrett01,Garrett07}, $A=50$ \cite{Lenzi01}, and $A=54$
\cite{Gadea06,Rudolph08} triplet nuclei. These studies have
suggested that the INC nuclear interaction in the $f_{7/2}$ shell
plays an important role in the explanation for the observed MED and
TED \cite{Bentley07,Zuker02}. In the upper $sd$ shell, however,
studies showed \cite{Vedora07} that important contributions to the
symmetry breaking come from the multipole Coulomb term and the
electromagnetic spin-orbit interaction, but not from the INC nuclear
interaction. Little has been explored for the upper $fp$-shell above
the $N=Z=28$ shell closure, and our knowledge on the
isospin-symmetry breaking in the mass-70 region is presently very
limited.

Recent advances in experiment have made it possible to collect very
exotic spectroscopic data. In the past few years, experimental
information on mirror nuclei of the upper $fp$-shell above the
doubly-magic nucleus $^{56}$Ni became available. The MED in the
$A\sim 60$ mass region were discussed \cite{Ekman05,Andersson05}. It
was suggested that the large MED in the mirror pair
$^{61}$Ga/$^{61}$Zn are due to the Coulomb monopole interaction and
the electromagnetic spin-orbit interaction. The positive-parity
high-spin states for the mirror pair $^{67}$As/$^{67}$Se were
observed \cite{Orlandi09}, and description of these states requires
inclusion of the $g_{9/2}$ orbit into the $fp$ model space. In Ref.
\cite{Kaneko10} we investigated the MED in $^{67}$As/$^{67}$Se using
the spherical shell model in the $pf_{5/2}g_{9/2}$ model space, and
suggested that the electromagnetic spin-orbit interaction and the
Coulomb monopole radial term are responsible for producing the large
MED at the high-spin states.

Quite recently, we have shown \cite{Kaneko12} that the experimental
Coulomb energy differences (CED) \cite{Angelis01,Nara07} between the
isospin $T = 1$ states in the odd-odd $N = Z$ nuclei ($T_z=0$) and
the analogue states in their even-even partners ($T_z=+1$) can be
accurately reproduced by shell-model calculations. The anomaly in
CED found in the pair $^{70}$Br/$^{70}$Se can be explained by the
electromagnetic spin-orbit interaction without the INC nuclear
interactions. Thus this work \cite{Kaneko12} and Refs.
\cite{Ekman05,Andersson05,Orlandi09,Kaneko10} seem to indicate that
in contrast to what has been suggested for the $f_{7/2}$-shell
nuclei, the important contribution to CED in the upper $fp$ shell
comes from the multipole Coulomb term and the electromagnetic
spin-orbit interaction, but not from the INC interaction.

However, we believe that the question whether the above conclusion
is general or not should be further investigated. In particular, it
is important to survey the entire mass region by using different
probes such as MED and TED. Unfortunately, for mirror and triplet
nuclei with $A>66$, there are no experimental data on MED and TED
available due to experimental difficulties with the $T=1, T_z=-1$
nuclei. Very recently, new observation of low-lying levels in
$^{66}$Se has been reported \cite{Obertelli11,Ruotsalainen13b},
which, with the data of $^{66}$As \cite{Ruotsalainen13a} and
$^{66}$Ge \cite{Stefanova03}, gives the experimental $A=66$ MED and
TED up to $J=6$. This is by now the heaviest triplet nuclei having
the TED data. It was shown \cite{Ruotsalainen13b} that the Coulomb
interaction alone can not account for the observed $A=$66 TED.

\begin{figure}[b]
\includegraphics[totalheight=7.5cm]{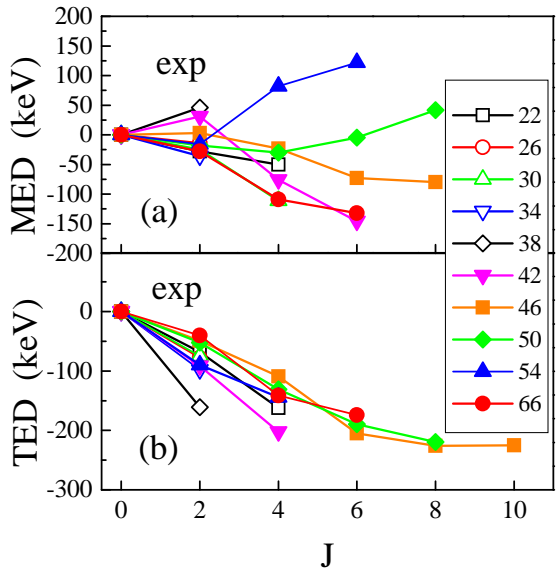}
  \caption{(Color online) Experimentally-known MED and TED for
  masses of $A=22-66$. Data are taken from Refs.
  \cite{Angelis01,Nara07,Obertelli11,Ruotsalainen13a,Stefanova03,
  Ruotsalainen13b,Hurst07,Ljungvall08,
  Lemasson12,McCutchan11,Fisher06,Uusitalo98,ENSDF}.}
  \label{fig1}
\end{figure}

The purpose of the present Rapid Communication is to investigate MED
and TED for the even-even nuclei with $A=66-78$ to extract
information on the isospin-symmetry breaking through detailed
analysis of the shell-model results. We show that, in addition to
the electromagnetic spin-orbit interactions, it is necessary to
involve the INC interaction to explain the new $A=66$ MED and TED
data. We discuss the type and strength of the INC force that is
phenomenologically added into the usual effective interaction. We
further show that the realistic nuclear interactions that contain
the isospin-breaking terms do not provide similar strengths for the
isotensor component that are needed to reproduce the $A=66$ MED and
TED data.

The MED in mirror-pair nuclei, defined by
\begin{equation}
 {\rm MED}({J}) = E_{x}(J,T=1,T_{z}=-1) - E_{x}(J,T=1,T_{z}=1),
  \label{eq:1}
\end{equation}
are regarded as measure of the charge-symmetry breaking in effective
nuclear interactions, which include the Coulomb force. In Eq.
(\ref{eq:1}), $E_{x}(J,T,T_{z})$ are the excitation energies of IAS
with spin $J$ and isospin $T$, distinguished by different $T_{z}$.
For $T=1$, the experimental MED for $A=22-66$ are shown in Fig.
\ref{fig1}(a). It is worth noting the behavior of the
$J^{\pi}=2^{+}$, $4^{+},$ and $6^{+}$ states in $A=42$ and $54$. For
each of these spin states, the two nuclei indicate nearly the same
magnitude in MED but opposite signs. This is called the
cross-conjugate symmetry in MED for the two extremes in the
$f_{7/2}$ shell \cite{Gadea06}.

The TED of $T=1$ states in triplet nuclei are defined by
\begin{eqnarray}
 {\rm TED}({J}) & = & E_{x}(J,T=1,T_{z}=-1) + E_{x}(J,T=1,T_{z}=1) \nonumber \\
& - & 2E_{x}(J,T=1,T_{z}=0),
  \label{eq:2}
\end{eqnarray}
which measures the charge-independence breaking. The experimental
TED for $A=22-66$ are shown in Fig. \ref{fig1}(b). The
characteristic feature is that all the TED have negative values and
exhibit similar spin-dependence. For example, the behavior of the
$A=46$ TED is almost identical to that of $A=50$ for the entire spin
range.
\begin{figure}[b]
\includegraphics[totalheight=7.0cm]{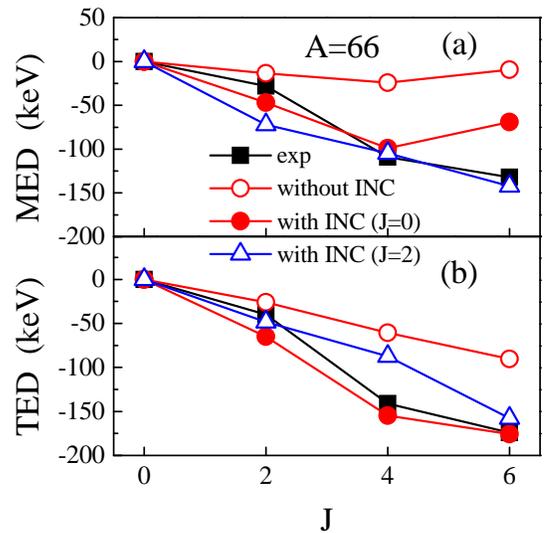}
  \caption{(Color online) Comparison of calculated MED and TED for $A=66$
  with experimental data taken from Refs.
  \cite{Obertelli11,Ruotsalainen13a,Stefanova03,Ruotsalainen13b}.}
  \label{fig2}
\end{figure}

We perform large-scale shell-model calculations in the
$pf_{5/2}g_{9/2}$ valence space. In the numerical calculations, the
recently-proposed SS method \cite{Mizusaki} is employed, which goes
beyond the usual Lanczos method and makes diagonalizations for the
current problem possible. We employ the modern JUN45 interaction
\cite{Honma09}, and add the multipole term $V_{CM}$ and monopole
term $\varepsilon_{ll}$ in the Coulomb interaction. In addition, the
single-particle energy shift $\varepsilon_{ls}$ due to the
electromagnetic spin-orbit interaction \cite{Nolen69} is also
included.

We first carry out a calculation with the same parameters taken from
our previous works \cite{Kaneko10,Kaneko12}, without considering the
INC interaction in the $pf_{5/2}g_{9/2}$ shell. In Fig. \ref{fig2},
the calculated MED and TED, indicated by open circles, are shown as
functions of spin $J$ for $A=$ 66. It is found that the magnitudes
of the obtained MED and TED are much smaller than the experimental
data although they indicate correct spin-dependent trends. Thus the
calculation shows that the multipole, monopole, and spin-orbit
forces are not enough to explain the data, and the INC force could
be important for reproducing the observed MED and TED.

We now add the INC interaction with the $J=0$ pairing terms
$V_{pp}=\beta_{pp}V^{J=0}_{pp}$, $V_{nn}=\beta_{nn}V^{J=0}_{nn}$,
and $V_{pn}=\beta_{pn}V^{J=0}_{pn}$ for all the orbits in our model
space, with $V^{J=0}_{pp}$, $V^{J=0}_{nn}$, and $V^{J=0}_{pn}$
being, respectively, the $pp$, $nn$, and $pn$ pairing interactions
for the matrix elements having a unit value (see Ref.
\cite{Kaneko13}). The strengths,
$\beta^{J=0}_{(2)}=\beta_{pp}+\beta_{nn}-2\beta_{pn}=100$ keV for
the isotensor and $\beta^{J=0}_{(1)}=\beta_{pp}-\beta_{nn}=300$ keV
for the isovector, are chosen so as to reproduce the experimental
CED, MED, and TED data for $A=66$ \cite{Ruotsalainen13b}. The
isotensor strength of 100 keV for $J=0$ is the same as that of the
empirical TED of the $A=42$ triplet \cite{Zuker02,Ruotsalainen13b}
in the $f_{7/2}$ shell.

Large differences have been found between the results with and
without the INC interaction. As one can see from Fig. \ref{fig2}(a),
the calculated MED with inclusion of INC can well reproduce the
experimental data \cite{Ruotsalainen13a} for $J=2$ and $J=4$, while
for $J=6$ the agreement is improved as compared to the calculation
without INC. The calculation shown in Fig. \ref{fig2}(b) reproduces
the TED data \cite{Obertelli11,Ruotsalainen13a} remarkably well. We
thus conclude that the INC interaction enhances the MED and TED
significantly, and is responsible for the isospin symmetry breaking
in the upper $fp$ shell. To reproduce the experimental MED for the
$6^{+}$ state, the $J=2$ INC pairing interaction alone may be used,
as discussed in Ref. \cite{Rudolph08,Davies13}. As seen in Fig.
\ref{fig2}, however, the $J=2$ INC pairing calculation with
$\beta^{J=2}_{(2)}=\beta^{J=2}_{(1)}=-200$ keV fails to describe the
MED for $J=2$ and TED for $J=4$, although it reproduces the MED data
for $J=6$. From Fig. \ref{fig2}(b), it is clear that the
experimental TED patterns are quite nicely reproduced by the $J=0$
INC pairing interaction \cite{Zuker02} alone.

\begin{figure}[b]
\includegraphics[totalheight=7.2cm]{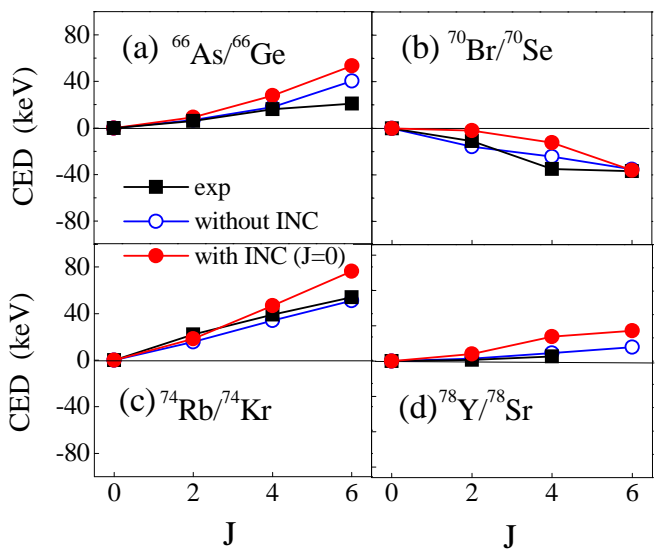}
  \caption{(Color online) Calculated CED for mass number $A=$66,
70, 74, and 78. Data are taken from Refs.
\cite{Obertelli11,Ruotsalainen13a,Stefanova03,Ruotsalainen13b}.}
  \label{fig3}
\end{figure}

We have carefully checked the proposed INC interaction to see
whether the present parameter choice of INC also reproduces CED (the
Coulomb energy difference between the $T=1$ states in the odd-odd
$N=Z$ nuclei and the analogue states in their even-even partners).
As seen in Fig. \ref{fig3}, inclusion of the same INC interaction
gives essentially a similar description for CED of $A=$ 66, 70, 74,
and 78, although the agreement with data is slightly better for the
results without INC \cite{Kaneko12}. We may thus conclude that the
present two-parameter INC interaction can correctly describe all the
existing data (MED, TED, and CED) for the mass $A=$ 66.

\begin{figure}[b]
\includegraphics[totalheight=7.2cm]{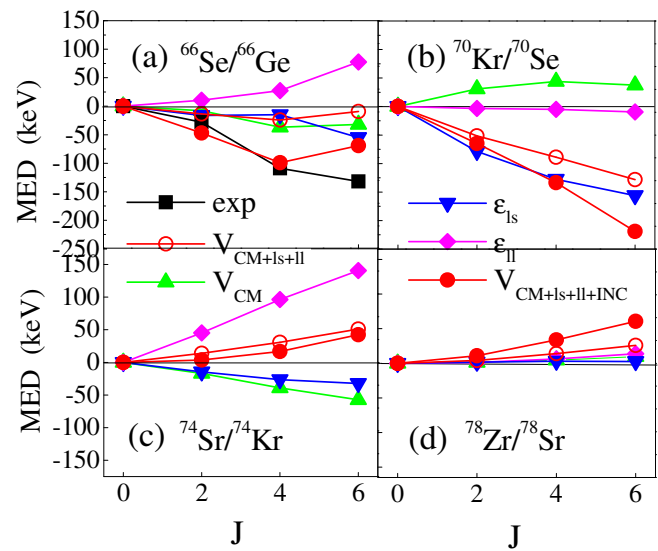}
  \caption{(Color online) Calculated MED for mass number $A=$66,
70, 74, and 78. Data are taken from Refs.
\cite{Obertelli11,Ruotsalainen13a,Stefanova03,Ruotsalainen13b}.}
  \label{fig4}
\end{figure}

With inclusion of the $J=0$ INC interaction, Fig. \ref{fig4} shows
the calculated MED for the upper $fp$-shell nuclei with $A=$ 66, 70,
74, and 78. Large variations in MED are seen in these nuclei, in
both the spin-dependent trend and the magnitude. We can analyze the
shell-model results by studying the components of the Hamiltonian.
In Fig. \ref{fig4}, the separated multipole, spin-orbit, and
monopole parts are denoted by $V_{CM}$, $\varepsilon_{ls}$, and
$\varepsilon_{ll}$, respectively. For the $A=66$ mirror pair
$^{66}$Se/$^{66}$Ge, $\varepsilon_{ls}$ and $V_{CM}$ have negative
values for the $J^{\pi}=0^{+}, 2^{+}$ and $4^{+}$ states, while
$\varepsilon_{ll}$ is positive. The net MED from summation of the
three terms are small and negative, which reproduces the
experimental MED for the $2^+$ state
\cite{Obertelli11,Ruotsalainen13a}. The MED components for the
$A=70$ mirror pair $^{70}$Kr/$^{70}$Se are large and negative for
$\varepsilon_{ls}$, but positive for $V_{CM}$ and nearly zero for
$\varepsilon_{ll}$. Since the magnitudes of $\varepsilon_{ls}$ are
larger, the total MED indicate negative values with large
magnitudes. This suggests that the spin-orbit contribution is
responsible for the negative MED in $^{70}$Kr/$^{70}$Se. For the
$A=74$ mirror pair $^{74}$Sr/$^{74}$Kr, the components indicate a
similar overall behavior as those in the $A=66$ pair. However,
$\varepsilon_{ll}$ is found significantly larger in the $A=74$ pair,
and is dominant in the summation. Therefore, the $A=74$ MED are
positive. Comparison of the corresponding components of $A=66$ and
74 suggests that the cross-conjugate \cite{Gadea06} feature of these
two pairs, when only the Coulomb part in the interaction is
considered, originates from the different contributions of the
monopole term $\varepsilon_{ll}$ in the Coulomb interaction. The INC
interaction tends to break the cross-conjugate symmetry. For
$^{78}$Zr/$^{78}$Sr, all components are small and positive, and the
total MED are therefore small and positive.

\begin{figure}[t]
\includegraphics[totalheight=7.2cm]{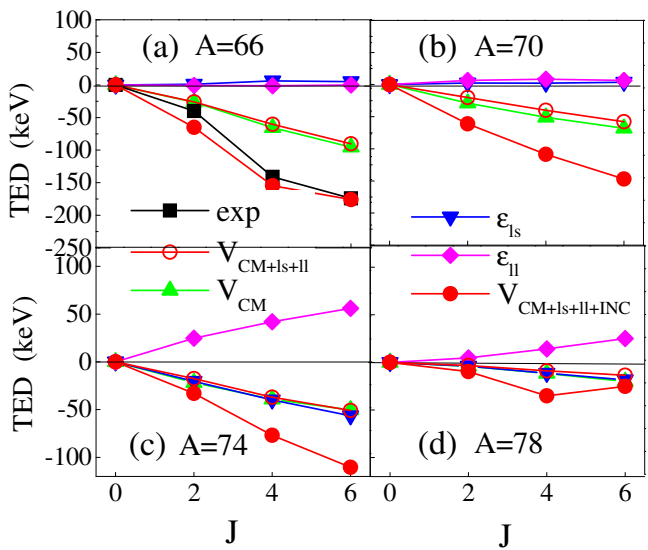}
  \caption{(Color online)
 Calculated TED for mass number $A=$66,
70, 74, and 78. Data are taken from Refs.
\cite{Obertelli11,Ruotsalainen13a,Stefanova03,Ruotsalainen13b}.}
  \label{fig5}
\end{figure}

Next, we discuss TED in the $T=1$ triplet nuclei. Figure \ref{fig5}
shows the calculated TED for $A=$ 66, 70, 74, and 78. One sees that,
in consistent with the known experimental TED for $A=22-66$ shown in
Fig. \ref{fig1}(b), all the total TED (denoted by
$V_{CM+ls+ll+INC}$) have negative values. We can easily see the
origin for the negative TED from its definition in Eq. (\ref{eq:2}).
This quantity is given by twice of the difference between the
average excitation energy of two even-even analogue states and that
of the odd-odd analogue state. In Fig. \ref{fig1}, the experimental
data indicate that the excitation energy of the odd-odd
$T=1,T_{z}=0$ state must be larger than either of the even-even
$T=1,T_{z}=\pm 1$ analogue states, and therefore the TED are
negative. This was explained \cite{OLeary99} in terms of the $T=1$
proton-neutron pairing. With increasing spin, the $pp$ and $nn$
pairs re-couple in the two even-even nuclei, while the $pn$ pairs
align in the odd-odd nuclei. The recoupling lowers the energy of the
proton-rich nuclei with $T_{z}=-1$ due to the Coulomb effect. We
have found that, although the Coulomb effect qualitatively accounts
for the negative TED, the components in Fig. \ref{fig5} show the
significant contribution from the INC interaction. It is clear that
without it, the calculated negative TED are less than half of the
experimental ones. This is consistent with the conclusion of
Ruotsalainen {\it et al.} (see Fig. 3(c) of Ref.
\cite{Ruotsalainen13b}) that the Coulomb (mutipole) interaction
alone does not account for the observed TED. As seen from Fig.
\ref{fig5}(a), the INC force enhances the TED magnitudes, which
explains well the experimental data for $A=66$. We thus conclude
that both the multipole component of the Coulomb force and the INC
force are responsible for the negative TED.

We have found that the observed TED in $A=66$ cannot be reproduced
by using different single-particle energies for protons and neutrons
alone. In other words, we do not see, at least for the $A=66$ case,
that the suggested INC effect can be effectively replaced by just
shifting single-particle energies between protons and neutrons. This
conclusion has been checked by using the single-particle energies
proposed by Trache {\it et al.} in Ref. \cite{Trache96} as an
example. The calculation tells us that there is essentially no
difference between the results using our single-particle energies
and those in Ref. \cite{Trache96}. However, the INC interaction is
necessary to reproduce the observed TED in $A=66$ no matter which
set of single-particle energies is used.

In Ref. \cite{Kaneko10}, some of the present authors performed MED
calculations for the odd-mass $A=67$ mirror pair nuclei. Now that
problem is recalculated with inclusion of the same INC forces used
in the present paper. It is found that the MED results presented in
Ref. \cite{Kaneko10} are further improved. The new MED for the
$3/2^-$ and $7/2^-$ states are $-45.2$ keV and $-30.8$ keV,
respectively, which are compared to the experimental ones $-43$ keV
and $-50$ keV. There is a clear improvement from our previous
results shown in Ref. \cite{Kaneko10} ($-15.4$ keV and $-72.2$ keV)
for which no INC forces were included. Furthermore, we confirm that
the MED of the positive-parity band built on the $9/2^+$ state are
also reproduced well by the new calculation. This is an additional
support to the current treatment of the INC forces in the upper $fp$
shell region.

$^{70}$Kr lies in the transitional region \cite{Hasegawa07} where
the shape-coexistence phenomenon has been discussed
\cite{Bouchez03,Fischer00,Gade05,Angelis12}. It is interesting to
examine the INC effects on the shape structure. The calculated
results predict an oblate shape for the yrast states and a weakly
prolate-deformation for the side band. By comparing the calculations
with and without the INC interaction, we find only little difference
in the results for both bands. We thus conclude that the INC force
does not affect the bulk property such as deformation, which is
consistent with the conclusion in Ref. \cite{Kaneko12,Nara07}.

Realistic interactions based on the modern charge-dependent
nucleon-nucleon ($NN$) forces, such as N$^{3}$LO, CD-Bonn, and AV18,
can well reproduce the $NN$ scattering data. The isotensor strengths
for the $NN$ forces have been calculated in the $G$ matrix formalism
\cite{Jensen95}, which, based on low-momentum interactions
$V_{low~k}$ \cite{Bogner03,Bogner07}, sums the particle-particle
ladders. The calculated strengths for $\beta^{J=0}_{(2)}$ are 225,
330, and 621 (all in keV) for N$^{3}$LO, CD-Bonn, and AV18,
respectively. These values are two to six times larger than the
strength adopted in the present work ($\beta^{J=0}_{(2)}=100$ keV).
Thus it is unlikely that the charge-dependent $NN$ forces will
provide the same description for MED and TED in the upper $fp$-shell
nuclei as the phenomenological treatment does. In the $f_{7/2}$
shell region, the observed $A=54$ TED were investigated using the
AV18 force. It was concluded \cite{Gadea06} that that force fails to
reproduce the experimental data.

To summarize, the recent experimental results
\cite{Obertelli11,Ruotsalainen13b} motivated us to carry out a
detailed shell-model analysis for MED and TED in the upper $fp$
shell region, aiming at a better understanding of isospin-symmetry
breaking in nuclear effective interactions. We have systematically
investigated MED and TED for the $T=1$, $T_z=0,\pm 1$ nuclei with
$A=66-78$ by performing large-scale shell-model calculations. Our
results for the upper $fp$-shell region have clearly shown that the
observed MED and TED for $A=66$ can only be explained by inclusion
of the INC nuclear interaction, in addition to the Coulomb force. We
have further predicted large, negative MED for $A=70$, which is
attributed to the INC interaction together with the same mechanism
leading to the anomalous CED between the isospin $T=1$ states in the
odd-odd $N=Z$ nucleus $^{70}$Br and the analogue states in its
even-even partner $^{70}$Se. It is also demonstrated that both the
INC interaction and the multipole Coulomb force determine the
negative TED systematically for $A=$ 66, 70, 74, and 78. We have
found that the isotensor strengths derived from the modern
charge-dependent forces (N$^{3}$LO, CD-Bonn, and AV18) deviate
strongly from the required values that reproduce the experimental
data. Presently it remains as an important open question why the
above-mentioned modern charge-dependent forces can not account for
the phenomenological strengths of the INC force. It would be
interesting to see if the three-nucleon forces based on the chiral
effective field theory \cite{chrial} can provide a solution.
Finally, experimental data for heavier $N<Z$ nuclei are much desired
to test our predictions and to understand the role of the INC force
in nuclear spectroscopy.

Research at SJTU was supported by the National Natural Science
Foundation of China (No. 11135005) and by the 973 Program of China
(No. 2013CB834401).



\end{document}